\documentclass[12pt]{article}
\usepackage{epsf,a4,latexsym}
\usepackage[small,bf]{caption}

\newcommand{\half}{\mbox{\small $\frac{1}{2}$}}          
\newcommand{\imp}{\mbox{\tiny $IMP$}}                    
\newcommand{\awi}{\mbox{\tiny $AWI$}}                    

\def\lsim{\mathrel{\rlap{\lower4pt\hbox{\hskip1pt$\sim$}}
    \raise1pt\hbox{$<$}}}                
\def\gsim{\mathrel{\rlap{\lower4pt\hbox{\hskip1pt$\sim$}}
    \raise1pt\hbox{$>$}}}                
\begin{document}

\title{
\vspace{-3.0cm}
\flushleft{\normalsize DESY 03-062} \\
\vspace{-0.35cm}
{\normalsize Edinburgh 2003/06} \\
\vspace{-0.35cm}
{\normalsize Leipzig LU-ITP 2003/011} \\
\vspace{-0.35cm}
{\normalsize Liverpool LTH 575} \\
\vspace{-0.35cm}
{\normalsize May 2003} \\
\vspace{0.5cm}
\centering{\Large \bf
       Non-perturbative renormalisation and improvement of the local vector
       current for quenched and unquenched Wilson fermions}}

\author{\large T. Bakeyev$^1$, M. G\"ockeler$^{2,3}$,
               R. Horsley$^4$, D. Pleiter$^5$,  \\
               P.~E.~L. Rakow$^6$, G. Schierholz$^{5,7}$
               and H. St\"uben$^{8}$\\[1em]
         -- QCDSF-UKQCD Collaboration -- \\[1em]
        \small 
          $^1$ Joint Institute for Nuclear Research,\\[-0.5em]
        \small
               RU-141980 Dubna, Russia\\[0.25em]
        \small
          $^2$ Institut f\"ur Theoretische Physik,\\[-0.5em]
        \small
                Universit\"at Leipzig, D-04109 Leipzig, Germany\\[0.25em]
        \small
          $^3$ Institut f\"ur Theoretische Physik,\\[-0.5em]
        \small
               Universit\"at Regensburg, D-93040 Regensburg, Germany\\[0.25em]
        \small
          $^4$ School of Physics,\\[-0.5em]
        \small
               University of Edinburgh, Edinburgh EH9 3JZ, UK\\[0.25em]
        \small
          $^5$ John von Neumann-Institut f\"ur Computing NIC,\\[-0.5em]
        \small
               D-15738 Zeuthen, Germany\\[0.25em]
        \small
          $^6$ Theoretical Physics Division,
               Department of Mathematical Sciences,\\[-0.5em]
        \small
               University of Liverpool, Liverpool L69 3BX, UK\\[0.25em]
        \small
          $^7$ Deutsches Elektronen-Synchrotron DESY,\\[-0.5em]
        \small
               D-22603 Hamburg, Germany\\[0.25em]
        \small
          $^8$ Konrad-Zuse-Zentrum f\"ur Informationstechnik Berlin,\\[-0.5em]
        \small
               D-14195 Berlin, Germany}

\date{October 29, 2003}

\maketitle

\vspace*{-0.25in}


\begin{abstract}
By considering the local vector current between nucleon states
and imposing charge conservation, we determine its
renormalisation constant and quark mass improvement coefficient
for Symanzik $O(a)$ improved Wilson fermions.
The computation is first performed for quenched fermions
(and for completeness also with unimproved fermions) and compared against
known results. The two-flavour unquenched case is then considered.
\end{abstract}

\clearpage


\section{Introduction}
\label{introduction}

A naive discretisation of fermions onto a hypercubical lattice
gives for the action errors of $O(a^2)$. However returning to
the continuum then leads to the famous fermion doubling problem
when we find $15$ extra copies of our original fermion.
To cure this Wilson \cite{wilson75a} added the `Wilson term' to the action
so that the copies decouple in the continuum limit:
but then discretisation errors are $O(a)$.
As the gluon part of the action (sum of plaquettes in this article)
has already $O(a^2)$ errors, it is desirable to also achieve this
for the fermion action. The Symanzik programme%
\footnote{For an introduction see, for example,
\cite{luscher98a, sommer97a, kronfeld02a}.}
allows a systematic reduction of errors%
\footnote{Ginsparg-Wilson fermions, an alternative formulation,
automatically have $O(a^2)$ errors \cite{niedermayer98a, capitani99a}
and have better chiral properties but are very CPU time consuming.}
to $O(a^2)$. An additional operator of dimension $5$
(the `clover term') is added to the Lagrangian with
a coefficient $c_{sw}$ suitably adjusted so that on-shell quantities
such as masses now have $O(a^2)$ errors.
For matrix elements it is also necessary to add further
higher dimensional operators to the original operator
to achieve $O(a)$ improvement. In this letter we shall
be concerned with the determination of the improvement
coefficients of the associated improvement operators
for the local vector current: $V_\mu^{(q)} \equiv \bar{q}\gamma_\mu q$. 
For this operator just two additional operators $am_q V^{(q)}_\mu$
and $ia\partial_\nu T^{(q)}_{\mu\nu}$ are required
giving the $O(a)$ improved vector current ${\cal V}_\mu^{(q)\imp}$,
and renormalised current ${\cal V}_\mu^{(q)R}$, as
\begin{equation}
  {\cal V}_\mu^{(q)R} = Z_V {\cal V}_\mu^{(q)\imp} \,,  \qquad
  {\cal V}_\mu^{(q)\imp} = ( 1 + am_q b_V )
      ( V^{(q)}_\mu + ia c_V \partial_\nu T^{(q)}_{\mu\nu}) \,,
                                                \nonumber
\end{equation}
with $T^{(q)}_{\mu\nu} = \bar{q} \sigma_{\mu\nu} q$,
$\sigma_{\mu\nu} = i [\gamma_\mu , \gamma_\nu] / 2$,  and
$\partial_\mu \phi(x) \equiv [ \phi(x+\hat{\mu}) - \phi(x-\hat{\mu})
] / (2a)$. The $ia\partial_\nu T^{(q)}_{\mu\nu}$ operator only
plays a role in non-forward matrix elements and will not be considered
further here. Thus to $O(a)$ improve the local vector current we need
to determine the improvement coefficient $b_V(g_0)$ (where $g_0$ is
the bare coupling constant).
Also as this current is not conserved on the lattice
then, as discussed in the next section,
it is renormalised with renormalisation constant $Z_V(g_0)$.
Perturbatively we have \cite{gabrielli91a,gockeler97a,sint97a,capitani00a}
to one loop (independently of the presence of fermions),
\begin{equation}
   Z_V(g_0) = 1 - 
         ( 0.174078 - 0.040069c_{sw} - 0.004586c_{sw}^2 ) g_0^2 + \ldots \,,
\label{pert_results_Z_V}
\end{equation}
where for unimproved fermions $c_{sw} = 0$ and for
$O(a)$ improved fermions $c_{sw} = 1 + O(g_0^2)$, together with
\begin{equation}
   b_V(g_0) = 1 + 0.15324 g_0^2 + \ldots \,.
\label{pert_results_b_V}
\end{equation}
However in regions where numerical simulations are performed,
$\beta \equiv 6/g_0^2 \sim 6.0-6.4$ for quenched fermions ($n_f=0$)
and $5.2-5.3$ for unquenched fermions ($n_f=2$), the above formulae
may not be applicable.

In this article we shall determine $Z_V$ and $b_V$ non-perturbatively
by considering (nucleon) matrix elements of the time component of the
local vector current and imposing charge conservation (the details
will be given in the following section). We shall first consider
quenched fermions and compare our results with known
results in the literature. (While most of the results will be for
$O(a)$ improved fermions, for comparison, we shall also briefly consider
the unimproved case, $c_{sw} = 0$.) This is then followed by
the unquenched case. Preliminary results have appeared in \cite{bakeyev02a}.


\section{The Conserved and Local Vector Currents}
\label{cvc+local}

There is an exact global symmetry of the lattice action
$q \to e^{-i\alpha_q}q$, $\bar{q} \to e^{i\alpha_q} \bar{q}$.
This global symmetry is flavour conservation (if you just rotate
the quarks of one flavour) or baryon number conservation (if you
rotate all quark flavours equally). Separate quark transformations
are possible because in pure QCD there are no flavour changing currents.
Upon using the Noether theorem this symmetry gives an exactly
conserved vector current or CVC of
\begin{equation}
   J^{(q)}_{\mu}( x + \half \hat{\mu})
      = \half \left[ \bar{q}_x(\gamma_\mu -1)U_\mu(x)q_{x+\hat{\mu}} -
                  \bar{q}_{x+\hat{\mu}}(\gamma_\mu +1)U_\mu(x)^\dagger q_x
                \right] \,.
\end{equation}
(Being conserved this current requires no renormalisation
constant and is $O(a)$ improved%
\footnote{This is of course only true when considering forward matrix elements.
For non-forward matrix elements the CVC requires the additional operator
$J_\mu^{(q)} \to {\cal J}_\mu^{(q)} \equiv J_\mu^{(q)} + 
\half ia c_{cvc} \half \left[ \partial_\nu T^{(q)}_{\mu\nu}(x) +
                      \partial_\nu T^{(q)}_{\mu\nu}(x+\hat{\mu})\right]$.}
.) By this we mean that the Ward Identity, WI, is
\begin{eqnarray}
   \langle \Omega \overline{\Delta}_\mu J^{(q)}_\mu(x + \half\hat{\mu})
   \rangle_{q,U} =
        \left\langle {\delta\Omega \over \delta q_x}q_x 
        \right\rangle_{q,U} + 
        \left\langle \bar{q}_x{\delta\Omega \over \delta \bar{q}_x} 
        \right\rangle_{q,U} \,,
\label{WI}
\end{eqnarray}
where $\Omega$ is an arbitrary functional of the $U$, $q$ and $\bar{q}$
fields. $\overline{\Delta}_\mu$ is the backward derivative,
$\overline{\Delta}_\mu \phi(x) \equiv [\phi(x) - \phi(x-\hat{\mu}]/a$.
(Although in eq.~(\ref{WI}) we integrate out the fermion fields,
$\langle \ldots \rangle_q$, and take the average over
the gauge fields, $\langle \ldots \rangle_U$,
the equation is already true configuration-by-configuration.)
We immediately see that if the region over which $\Omega$ is defined
does not contain $x$ the RHS of eq.~(\ref{WI}) vanishes so that
$\overline{\Delta}_\mu J^{(q)}_\mu(x+\half\hat{\mu}) = 0$.
If $\Omega$ contains $x$ then the RHS effectively `counts' the
number of quarks and anti-quarks in $\Omega$.

Here, in this study, we shall take $\Omega \to B(t) \bar{B}(0)$ 
where $B$ is the standard (stationary, or $\vec{p}=0$) nucleon operator,
containing two $u$ quarks and one $d$ quark summed over the spatial planes%
\footnote{We actually take
$\Gamma^{unpol}_{\alpha\beta} B_\beta(t) \bar{B}_\alpha(0)$
where $\Gamma^{unpol} \equiv \half (1 + \gamma_4)$
projects out the unpolarised component of the nucleon field,
$B_\alpha(t) = \sum_{\vec{x}} \epsilon^{ijk} u_\alpha^i(\vec{x},t) 
[ u^j(\vec{x},t)^T C\gamma_5d^k(\vec{x},t)]$.}.
(This was previously used as part of a project to determine moments
of structure functions; other choices are of course possible.)
Inserting this $\Omega$ in eq.~(\ref{WI}) and also summing over 
spatial points of $x \equiv (\vec{x},\tau)$ gives
\begin{equation}
   R[J^{(q)}_4(\tau)] - R[J^{(q)}_4(\tau -1)]
      = - \chi^{(q)} \left( \delta_{\tau , t} - \delta_{\tau , 0} \right) \,,
\label{Rdiff}
\end{equation}
where $R$ is defined as the ratio of three-point to two-point 
correlation functions,
\begin{equation}
   R[{\cal O}(\tau)]
      \equiv { \langle B(t) {\cal O}(\tau) \bar{B}(0) \rangle_{q,U}
                     \over \langle B(t) \bar{B}(0) \rangle_{q,U}} \,,
\label{Rdef}
\end{equation}
and $\chi^{(u)} = 2$, $\chi^{(d)} = 1$. With
\begin{equation}
   R[J^{(q)}_4(\tau)] =
       \left\{ \begin{array}{cl}
                c^{(q)}_1  & 0 \le \tau < t       \\
                c^{(q)}_2  & t \le \tau \le N_T-1 \\
             \end{array}
       \right. \,,
\end{equation}
for an $N_S^3 \times N_T$ lattice with $c_i^{(q)}$ ($i=1, 2$) being constant
eq.~(\ref{Rdiff}) can be solved to give the result
\begin{equation}
   c^{(q)}_1 - c^{(q)}_2 = \chi^{(q)} \,.
\label{jump}
\end{equation}
So $R$ should be a constant with jump, or discontinuity,
given by $\chi^{(q)}$. Note that this should be true to `machine accuracy'.
(Indeed in this special case $R$ may also be taken to be the ratio of three-
to two-point correlators for a single configuration.)
This result may also be shown using transfer matrix methods,
as indicated in the appendix, where it is also demonstrated that
$c_2^{(q)} \ll c_1^{(q)}$.

\begin{figure}[t]
   \hspace*{0.50in}
   \begin{tabular}{cc}
      \epsfxsize=5.00cm \epsfbox{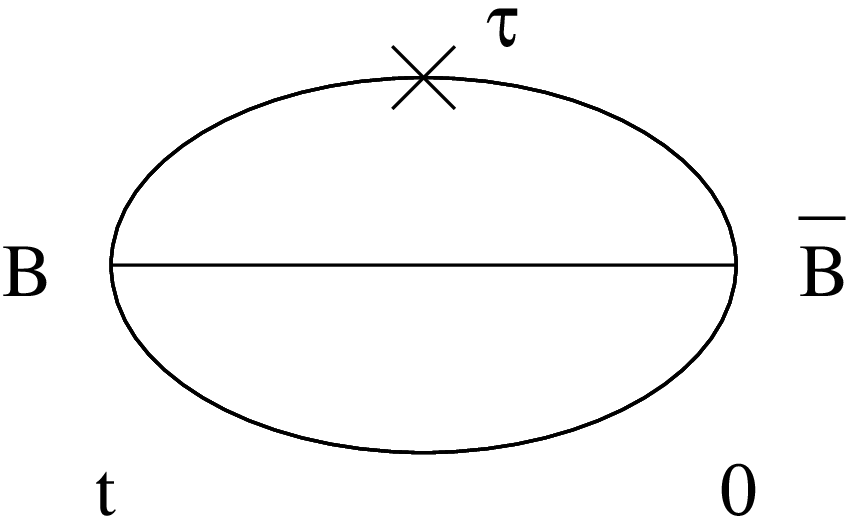}     &
      \hspace{1.0cm}
      \epsfxsize=5.00cm \epsfbox{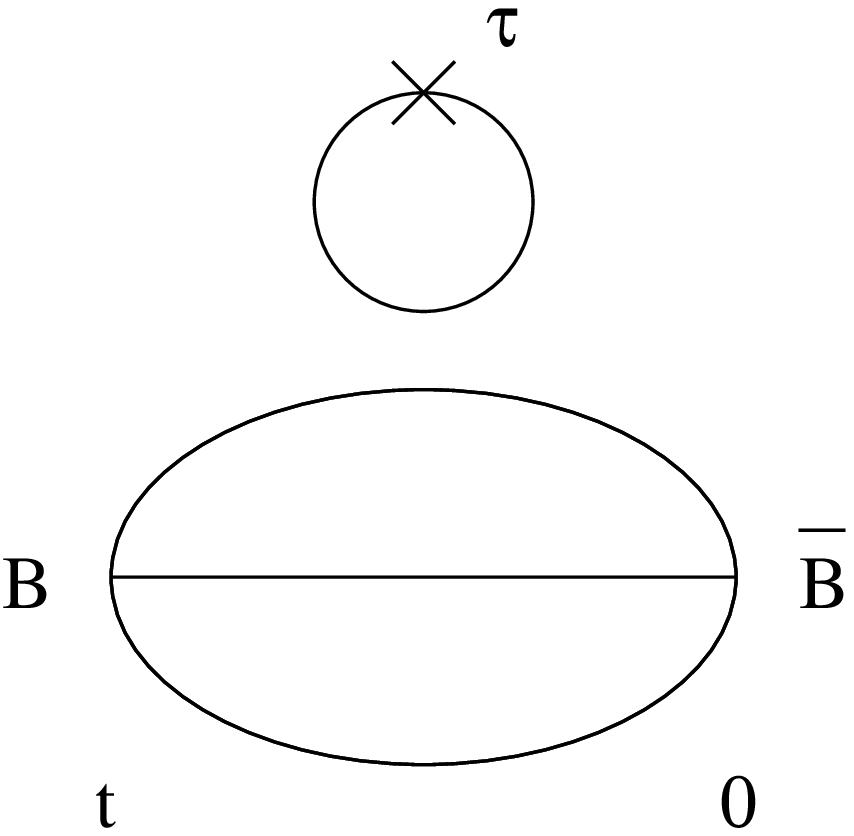}
   \end{tabular}
   \caption{The quark-line connected diagram, left hand picture,
            and quark-line disconnected diagram, right hand picture.
            The cross denotes the operator insertion ${\cal O}(\tau)$.}
   \label{fig_3pt_conn+disconn}
\end{figure}

\begin{figure}[t]
   \hspace*{0.35in}
   \epsfxsize=12.00cm \epsfbox{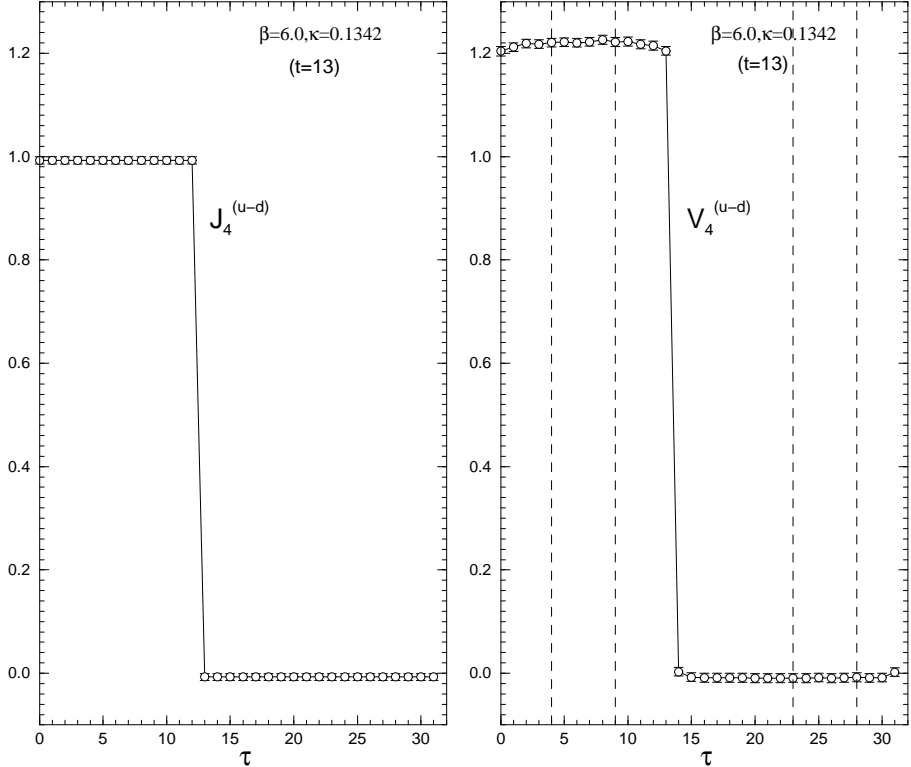}
   \caption{$R[J_4^{(u-d)}(\tau)]$ and
            $R[V_4^{(u-d)}(\tau)]$ plotted against the operator
            position $\tau$ for
            the quenched ($n_f=0$) data set $\beta = 6.0$,
            $\kappa = 0.1342$ on an $N_S^3\times N_T = 16^3\times 32$
            lattice with $t=13$. Typical fit intervals for
            $V_4^{(u-d)}$ are given by the pairs of vertical
            dashed lines. The relationship between $am_q$ and
            $\kappa$ is given in eq.~(\ref{defofmq}).}
   \label{fig_Rratio_bare_b6p00kp1342_standard}
\end{figure}

The lattice computation of the three- and two-point functions for
$R[{\cal O}(\tau)]$ follows the standard way, see eg \cite{gockeler95a}.
(The source $\bar{B}(0)$ and sink $B(t)$ have been additionally
improved by non-relativistic projection and Jacobi smearing
to increase the overlap with the ground state nucleon,
as described for example in \cite{gockeler94a}. This does not affect
the arguments given above.)
We must, in principle, compute a quark-line connected contribution
(to the operator) and a quark-line disconnected term as shown in
Fig.~\ref{fig_3pt_conn+disconn}.
This latter term is numerically extremely difficult to compute,
due to ultra-violet fluctuations. However for the CVC this
term is in fact constant. This may be easily seen by substituting
$\Omega \to \langle B(t) \bar{B}(0) \rangle_q$ into the  
WI, eq.~(\ref{WI}). The RHS is then zero and
so using the same argument as before the appropriate ratio
is constant for all $\tau$. There is thus no contribution to the
discontinuity. Indeed on a finite lattice, we would expect this constant
to be exponentially small (with exponent proportional to $N_T$).
Physically there is no quark-line disconnected term because creating a
quark-antiquark pair cannot change the charge. Nevertheless with an
eye on the computation of the local current it is useful to consider
the difference between the $u$ and $d$ operators,
ie the non-singlet operator,
\begin{equation}
   {\cal O}^{(u-d)} \equiv {\cal O}^{(u)} - {\cal O}^{(d)} \,,
\end{equation}
in which the quark-line-disconnected terms cancel and so we find
\begin{eqnarray}
   R[J^{(u-d)}_4(\tau)] = \left\{ \begin{array}{ll}
                                     c_1^{(u-d)} & 0 \le \tau \le t  \\
                                     c_2^{(u-d)} & t \le \tau \le N_T-1
                                  \end{array} \right. \,,
\label{RNS}
\end{eqnarray}
where the discontinuity between the two constants, $\Delta R [J^{(u-d)}_4]$,
in eq.~(\ref{RNS}) is given by
\begin{equation}
   \Delta R [J^{(u-d)}_4] \equiv c_1^{(u-d)} - c_2^{(u-d)}
                          = \chi^{(u)} - \chi^{(d)} = 1 \,.
\end{equation}
An example for the ratio $R[J^{(u-d)}_4]$ is shown
in Fig.~\ref{fig_Rratio_bare_b6p00kp1342_standard}.
For $J_4^{(u-d)}$ a very good signal is observed.
For the jump, as expected, we find $1$ to within machine precision.

The local vector current (LVC) does not obey the WI
given in eq.~(\ref{WI}), but as $J_\mu^{(q)} = V^{(q)}_\mu + O(a)$
there is an additional term formally of $O(a)$ on the RHS
of this equation. However perturbatively expanding
eq.~(\ref{WI}) gives loop graphs with ultra-violet divergences
$\sim 1/a$ and so this additional term gives a contribution.
Thus to obey the WI to $O(a^2)$, $V_\mu$ must be renormalised
with renormalisation constant $Z_V$ and the quark-line disconnected
term in Fig.~\ref{fig_3pt_conn+disconn} may give an $O(a)$ contribution.
Again to avoid computing this term we consider the non-singlet
operator%
\footnote{Tests from computing $R[V_4^{(q)}(\tau)]$ for $q = u$, $d$
separately showed indirectly that the contribution from the quark-line
disconnected term must numerically be very small. So the difference
between the singlet and non-singlet renormalisation constant
must also be very small.}.
The necessity for a renormalisation constant is also reflected
in the fact that $\Delta R[V_4^{(u-d)}]$ will not be equal to one.
This is illustrated in the RH picture
of Fig.~\ref{fig_Rratio_bare_b6p00kp1342_standard}.
So we can {\it define} the renormalisation and improvement constants
($Z_V$ and $b_V$ respectively) by demanding that the local current
has the same behaviour as the conserved current, ie%
\footnote{It is more precise to define
\begin{eqnarray}
   Z_V &=& \lim_{am_q \to 0} ( \Delta R[ V_4^{(u-d)} ] )^{-1} \,,
                                                \nonumber  \\
   b_V &=& \lim_{am_q \to 0} {\partial \over \partial(am_q)}
                             \ln ( \Delta R[ V_4^{(u-d)} ] )^{-1} \,.
                                                \nonumber
\end{eqnarray}.}
\begin{eqnarray}
   Z_V ( 1 + am_q b_V) \equiv
        \left( \Delta R[ V_4^{(u-d)} ] \right)^{-1} \,.
\label{defZvbv}
\end{eqnarray}
Thus upon plotting the data for $( \Delta R[ V_4^{(u-d)} ] )^{-1}$
against $am_q$, we see that the intercept gives $Z_V$ while the gradient
gives $Z_Vb_V$. Note that $Z_V$ has potential $O(a^2)$ differences
to other definitions of the renormalisation constant while $b_V$
has possible $O(a)$ differences.

While this procedure is correct in the quenched case, for the unquenched
case for complete $O(a)$ cancellation, $g_0$ in $Z_V$ should be replaced
by $\tilde{g}_0$ where we have $\tilde{g}_0^2 = g_0^2(1 + b_g am_q)$,
\cite{luscher96a}. Thus the LHS of eq.~(\ref{defZvbv}) should now become
$Z_V ( 1 + am_q (b_V + \half b_g {g_0 \over Z_V} 
               {\partial Z_V \over \partial g_0}))$.
So while the intercept still gives $Z_V$, the gradient and hence $b_V$
is modified. $b_g$ is only known to one loop perturbation theory,
$b_g = 0.01200 n_f g_0^2 + O(g_0^4)$, \cite{luscher96a}.
Estimating $\partial Z_V / \partial g_0$ for unquenched fermions from
the Pad{\'e} fit, eq.~(\ref{padefit})
and Table~\ref{table_padefit} to be $\sim -1.3$ gives roughly for
the additonal term a decrease of $\sim 0.02$.
As numerically $b_V$ will turn out to be $\sim 2$,
then this could give a $1-2\%$ correction. At present, due to 
uncertainties in estimating this extra term, we shall ignore
this small correction factor.


\section{Simulation parameters and raw results for $O(a)$ improved fermions}

We have made runs at $\beta = 6.0$, $6.2$
and $6.4$ for quenched fermions and
$\beta = 5.20$, $5.25$ and $5.29$ for unquenched fermions.
The run parameters and results for 
$(\Delta R(V_4^{(u-d)}])^{-1}$
are given in Tables~\ref{table_quenched_run_params}
and \ref{table_unquenched_data_sets}.
\begin{small}
\begin{table}[t]
   \begin{center}
      \begin{tabular}{||l|l|l|l|c||c||}
         \hline
\multicolumn{1}{||c}{$\beta$}  &
\multicolumn{1}{|c}{$c_{sw}$}  &
\multicolumn{1}{|c}{$\kappa$}  & 
\multicolumn{1}{|c}{$N_S^3\times N_T$}  &
\multicolumn{1}{|c||}{$\#$ configs.} &
\multicolumn{1}{c||}{$(\Delta R[V_4^{(u-d)}])^{-1}$}  \\
         \hline
6.0 & 1.769 & 0.1320 & $16^3\times 32$ & $O(450)$  & 0.8839(4) \\
6.0 & 1.769 & 0.1324 & $16^3\times 32$ & $O(560)$  & 0.8717(3) \\
6.0 & 1.769 & 0.1333 & $16^3\times 32$ & $O(560)$  & 0.8418(4) \\
6.0 & 1.769 & 0.1338 & $16^3\times 32$ & $O(520)$  & 0.8240(6) \\
6.0 & 1.769 & 0.1342 & $16^3\times 32$ & $O(740)$  & 0.8124(8) \\
         \hline
         \hline
6.2 & 1.614 & 0.1333 & $24^3\times 48$ & $O(300)$  & 0.8691(2) \\
6.2 & 1.614 & 0.1339 & $24^3\times 48$ & $O(300)$  & 0.8503(2) \\
6.2 & 1.614 & 0.1344 & $24^3\times 48$ & $O(300)$  & 0.8342(2) \\
6.2 & 1.614 & 0.1349 & $24^3\times 48$ & $O(470)$  & 0.8186(2) \\
         \hline
         \hline
6.4 & 1.526 & 0.1338 & $32^3\times 48$ & $O(220)$  & 0.8624(1) \\
6.4 & 1.526 & 0.1342 & $32^3\times 48$ & $O(120)$  & 0.8504(2) \\
6.4 & 1.526 & 0.1346 & $32^3\times 48$ & $O(220)$  & 0.8376(1) \\
6.4 & 1.526 & 0.1350 & $32^3\times 48$ & $O(320)$  & 0.8253(1) \\
6.4 & 1.526 & 0.1353 & $32^3\times 64$ & $O(260)$  & 0.8163(3) \\
         \hline
      \end{tabular}
   \end{center}
\caption{Parameter values used in the quenched simulations
         with improved fermions, together with
         $(\Delta R[V_4^{(u-d)}])^{-1}$.
         The fit intervals chosen were
         $[n_l, n_u]$ and $[N_T-n_l, N_T-n_u]$ where
         $[n_l, n_u] = [9, 13]$, $[6,11]$, $[7,16]$ for
         $\beta = 6.0$, $6.2$ and $6.4$ respectively.}
\label{table_quenched_run_params}
\end{table}
\end{small}
\begin{small}
\begin{table}[t]
   \begin{center}
      \begin{tabular}{||l|l|l|l|c|l||c||}
         \hline
\multicolumn{1}{||c}{$\beta$}          &
\multicolumn{1}{|c}{$c_{sw}$}          &
\multicolumn{1}{|c}{$\kappa_{sea}$}    & 
\multicolumn{1}{|c}{$N_S^3\times N_T$} &
\multicolumn{1}{|c}{$\#$ trajs.}       &
\multicolumn{1}{|c||}{Group} &
\multicolumn{1}{c||}{$(\Delta R[V_4^{(u-d)}])^{-1}$}  \\
         \hline
 5.20 & 2.0171 & 0.1342 & $16^3\times 32$ & 5000 & QCDSF & 0.8060(09)\\
 5.20 & 2.0171 & 0.1350 & $16^3\times 32$ & 8000 & UKQCD & 0.7731(10)\\
 5.20 & 2.0171 & 0.1355 & $16^3\times 32$ & 8000 & UKQCD & 0.7537(16)\\
         \hline
 5.25 & 1.9603 & 0.1346 & $16^3\times 32$ & 2000 & QCDSF & 0.8019(08)\\
 5.25 & 1.9603 & 0.1352 & $16^3\times 32$ & 8000 & UKQCD & 0.7781(08)\\
 5.25 & 1.9603 & 0.13575& $24^3\times 48$ & 2000 & QCDSF & 0.7560(06)\\
         \hline 
 5.29 & 1.9192 & 0.1340 & $16^3\times 32$ & 4000 & UKQCD & 0.8328(07)\\
 5.29 & 1.9192 & 0.1350 & $16^3\times 32$ & 5000 & QCDSF & 0.7948(03)\\
 5.29 & 1.9192 & 0.1355 & $24^3\times 48$ & 2000 & QCDSF & 0.7747(04)\\
         \hline
      \end{tabular}
   \end{center}
   \caption{Data sets used in the unquenched,
            $n_f=2$, simulations together with
            $(\Delta R[V_4^{(u-d)}])^{-1}$. The fit intervals chosen were
            $[9,13]$ and $[N_T-9,N_T-13]$.}
   \label{table_unquenched_data_sets}
\end{table}
\end{small}

The bare quark mass is defined as
\begin{equation}
   am_q(\kappa, g_0) 
          = {1\over 2} \left( {1 \over \kappa} - {1 \over \kappa_c(g_0)}
                       \right) \,,
\label{defofmq}
\end{equation}
where $\kappa$ is the hopping parameter of the simulation. It is necessary
to find $\kappa_c(g_0)$, ie the critical point where the (bare) quark
mass vanishes. From PCAC we know that the quark mass is $\propto m_{ps}^2$
and we can use this to determine where the quark mass vanishes.
However a more precise/stable determination was often possible if the
PCAC quark mass $m_q^{\awi}$ was used, so we fitted
the dimensionless quantity, 
\begin{eqnarray}
   X &=& F( r_0 m_q^{\imp} )
                                                  \nonumber \\
     &\equiv& B_1 r_0 m^{\imp}_q  + B_2 (r_0 m^{\imp}_q)^2 + \ldots \,,
\label{pcac} 
\end{eqnarray}
where $X = r_0 m_q^{\awi}$ and $m^{\imp}_q = m_q ( 1 + b_m am_q )$
is the $O(a)$ improved quark mass. However determining $am_q^{\awi}$
requires a knowledge of the $c_A$ improvement coefficent. While this
is known for quenched fermions \cite{luscher96b} for unquenched
fermions this more sensitively affects the value of $am_q^{\awi}$.
Thus in this case we have set $X = (r_0 m_{ps})^2$.
$r_0$ in eq.~(\ref{pcac}) is the `force' scale, \cite{sommer93a}.
While $r_0$ could be omitted for quenched fermions, for unquenched fermions
$r_0$ becomes quark mass as well as coupling constant dependent
and we use the results given in \cite{booth01a},
supplemented by \cite{irving03a}.
For quenched fermions, we fitted both $B_1$ and $B_2$ coefficients,
which practically meant that no $b_m$ coefficient was needed,
being absorbed into the $B_2$ coefficient, while for unquenched fermions,
as we had only three quark mass values we set $B_2 = 0$ and
used a tadpole improved value of $b_m$, following the prescription
given in \cite{capitani00a} (and using the plaquette values given in
\cite{booth01a}).
This gave the results for $\kappa_c$ in Table~\ref{table_kappa_c}.
\begin{table}[t]
   \begin{center}
      \begin{tabular}{||l||l||l|l|l||}
         \hline
\multicolumn{1}{||c||}{$\beta$}   &
\multicolumn{1}{c||}{$\kappa_c$} &
\multicolumn{1}{c|}{$Z_V$}      &
\multicolumn{1}{c|}{$Z_Vb_V$}   &
\multicolumn{1}{c||}{$b_V$}     \\
         \hline
    6.0  & 0.135201(9)    & 0.7799(7)  & 1.168(10)  & 1.497(13) \\
    6.2  & 0.135803(3)    & 0.7907(3)  & 1.135(06)  & 1.436(08) \\
    6.4  & 0.135744(1)    & 0.8027(2)  & 1.116(04)  & 1.391(05) \\
         \hline
         \hline
    5.20 & 0.136072(10)   & 0.7304(18) & 1.472(44)  & 2.015(61) \\
    5.25 & 0.136287(09)   & 0.7349(09) & 1.460(31)  & 1.987(43) \\
    5.29 & 0.136368(09)   & 0.7420(07) & 1.411(19)  & 1.902(25) \\
         \hline
      \end{tabular}
   \end{center}
   \caption{$\kappa_c(g_0)$, the intercept $Z_V$,
            gradient $Z_Vb_V$ and $b_V$ for the quenched data sets
            ($\beta = 6.0$, $6.2$ and $6.4$) and for the unquenched
            data sets ($\beta = 5.20$, $5.25$ and $5.29$).}
   \label{table_kappa_c}
\end{table}
For quenched fermions these numbers are in good agreement with those
given in \cite{luscher96b}.

\begin{figure}[t]
   \hspace*{0.35in}
   \epsfxsize=12.00cm 
      \epsfbox{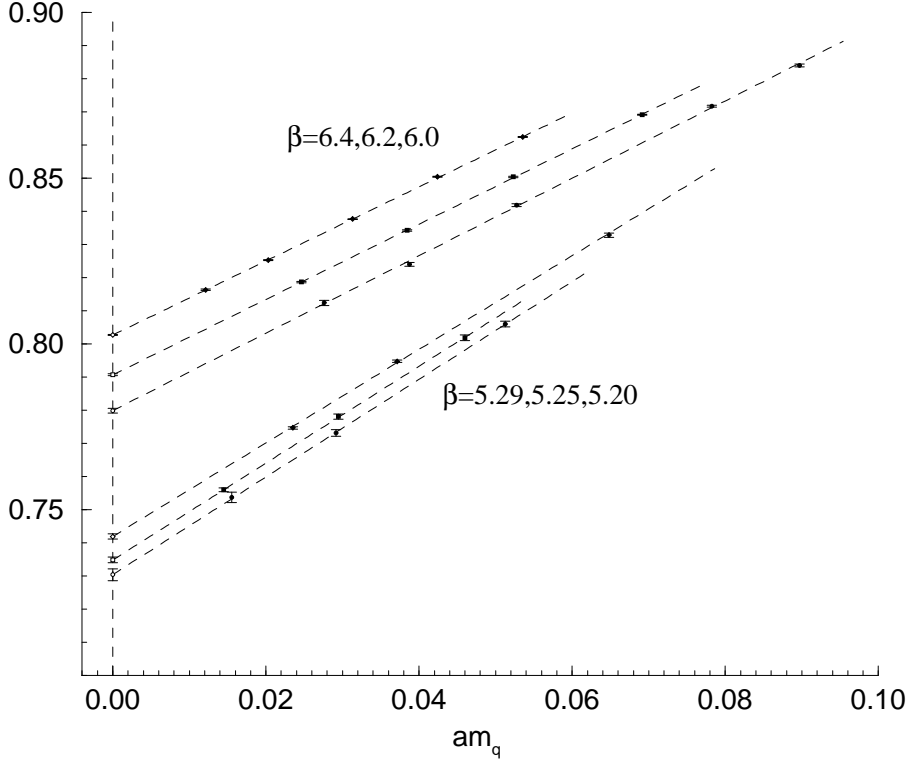}
   \caption{Linear extrapolations for $(\Delta R[V_4^{(u-d)}])^{-1}$
            for the quenched data at $\beta = 6.0$, $6.2$ and $6.4$
            (upper set of lines, from $\beta = 6.4$, upper line,
            to $6.0$, lower line) and for the unquenched data
            at $\beta = 5.20$, $5.25$ and $5.29$ (lower set of
            lines, with the highest being for $\beta = 5.29$).}
   \label{fig_V_0_1u-1d.p0_020509_1244_lat02_expt_wrtup_Zvpap}
\end{figure}

As discussed in section~\ref{cvc+local} we now measure the intercept and
gradient of $(\Delta R[V_4^{(u-d)}])^{-1}$ against $am_q$. The results
are shown in Fig.~\ref{fig_V_0_1u-1d.p0_020509_1244_lat02_expt_wrtup_Zvpap}
for both the quenched and unquenched data.
Very good linearity is seen in the results,
which enables a precise estimation of the intercept and slope.
The results are also given in Table~\ref{table_kappa_c}.
While these extrapolations have no problems the determination of
$\kappa_c$ may be more problematic.
To give an estimate of possible further errors, if $\kappa_c$
is varied by $\delta \kappa_c$, then we have
\begin{equation}
   \delta Z_V \sim b_V Z_V \delta (am_q) \,, \quad
   \delta b_V \sim b_V^2 \delta (am_q) \,,
   \quad \mbox{with} \quad
   \delta (am_q) \sim {1\over 2 \kappa_c^2} \delta \kappa_c  \,.
\end{equation}
For example, taking $\delta\kappa_c \sim 0.2 \times 10^{-4}$ gives
$\delta (am_q) \sim 0.0005$ and $\delta Z_V \sim 0.0006$,
$\delta b_V \sim 0.001$. These are of the same order or
less than the given statistical error.


\section{$Z_V$ and $b_V$ for $O(a)$ improved fermions}

We are now in a position to give our final results for $Z_V$ and $b_V$.
From eq.~(\ref{defZvbv}), dividing the gradient in
Fig.~\ref{fig_V_0_1u-1d.p0_020509_1244_lat02_expt_wrtup_Zvpap}
by the intercept gives in Table~\ref{table_kappa_c}
in the third and fifth columns $Z_V$ and $b_V$ respectively.

As for both quenched and unquenched fermions we have
three $\beta$ values, we can attempt to make a Pad{\'e}-type fit
of the form 
\begin{equation}
   P(g_0) = { 1 + p_1 g_0^2 + p_2 g_0^4 \over
               1 + q_1 g_0^2 } \,,
\label{padefit}
\end{equation}
for $Z_V(g_0)$, $b_V(g_0)$ constrained to reproduce the weak coupling
results, eqs.~(\ref{pert_results_Z_V}), (\ref{pert_results_b_V}).
There are thus $2$ free parameters. In Table~\ref{table_padefit}
\begin{table}[t]
   \begin{center}
      \begin{tabular}{||c||c|c|c||}
         \hline
         \hline
\multicolumn{1}{||c||}{}    &
\multicolumn{1}{c|}{$p_1$}  &
\multicolumn{1}{c|}{$p_2$}  &
\multicolumn{1}{c||}{$q_1$} \\
         \hline
\multicolumn{4}{||c||}{Quenched ($n_f=0$)} \\
         \hline
    $Z_V$  & -0.634   &  0.0196  & -0.504 \\
    $b_V$  & -0.627   & -0.0444  & -0.781 \\
         \hline
         \hline
\multicolumn{4}{||c||}{Unquenched ($n_f=2$)} \\
         \hline
    $Z_V$  & -0.796   &  0.0652  & -0.667   \\
    $b_V$  & -0.614   & -0.0432  & -0.767   \\
         \hline
         \hline
      \end{tabular}
   \end{center}
   \caption{The Pad{\'e} fit results as defined in eq.~(\ref{padefit}).}
   \label{table_padefit}
\end{table}
we give the results of the fits. Note however that even though
we have matched to the weak coupling results, as the range of $\beta$
where the numerical results lie is rather small and far away from this
region, intermediate regions may not be represented so well.
We estimate total errors on these Pad{\'e} results to be about $\half\%$
for $Z_V$ for both the quenched and unquenched cases and for $b_V$,
$1\%$ and $2\%$ for quenched and unquenched fermions respectively.
(Note also that for the unquenched results there is a further error
in $b_V$ of $1 - 2\%$ due to an uncertainty in $b_g$ as discussed
previously in section~\ref{cvc+local}.)

\begin{figure}[t]
   \hspace*{0.35in}
   \epsfxsize=12.00cm 
      \epsfbox{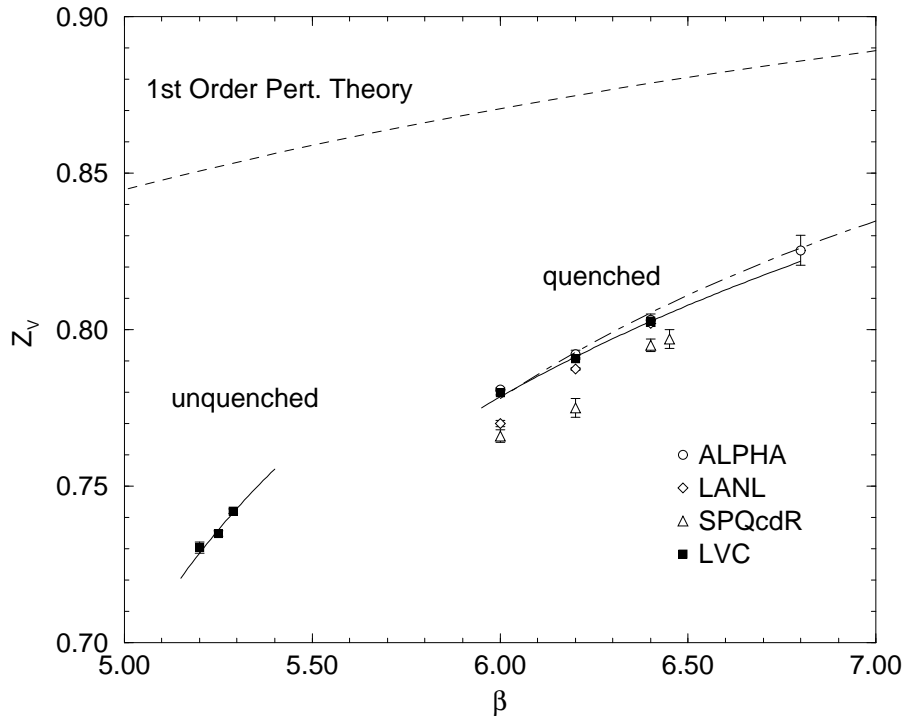}
   \caption{$Z_V$ (LVC, filled squares) determined in
            this work for quenched and unquenched $O(a)$ improved
            fermions. For quenched fermions a comparison is made
            with ALPHA, \protect{\cite{luscher96c}},
            LANL, \protect{\cite{bhattachary01a}} and
            SPQcdR, \protect{\cite{becirevic02a}}.  Pad{\'e}
            fits are also given for our (full lines) and the ALPHA 
            (dot-dashed line) results. The dashed line is the
            first order perturbative result, eq.~(\ref{pert_results_Z_V}).}
   \label{fig_zv+bv_zv_beta_magic_qu+dyn_zvpap}
\end{figure}

\begin{figure}[t]
   \hspace*{0.35in}
   \epsfxsize=12.00cm 
      \epsfbox{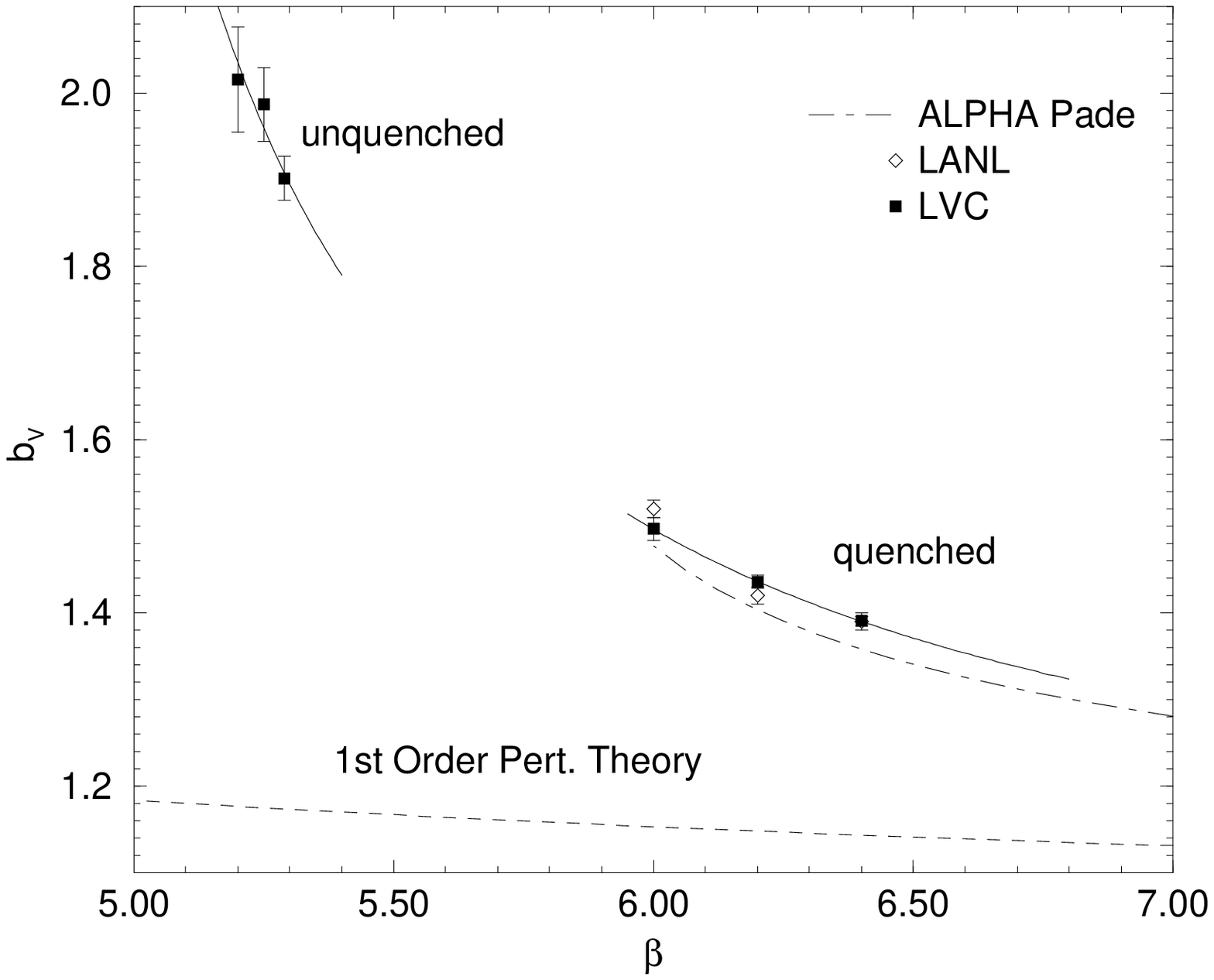}
   \caption{$b_V$ (LVC, filled squares) determined in
            this work for both quenched and unquenched $O(a)$
            improved fermions. Also shown are our Pad{\'e} fits 
            (full lines) and that from ALPHA, \protect{\cite{luscher96c}}
            (dot-dashed line), and the LANL, \protect{\cite{bhattachary01a}}
            results for quenched fermions. The dashed line is the
            first order perturbative result, eq.~(\ref{pert_results_b_V}).}
   \label{fig_zv+bv_bv_beta_magic_qu+dyn_zvpap}
\end{figure}

Alternative non-perturbative determinations for quenched
$O(a)$ improved fer\-mions have been given
by the ALPHA Collaboration, using the Schr\"od\-inger functional
method, \cite{luscher96c}, the LANL Collaboration,
\cite{bhattachary00a,bhattachary01a} using other
Ward identities and the SPQcdR Collaboration \cite{becirevic02a}
using the RI-MOM scheme, \cite{martinelli94a}.
All these other methods have reasonable agreement with the results
given here, as can be seen in the following pictures.
(One should remember that $Z_V$ definitions
in particular RI-MOM can vary by at least $O(a^2)$ and $b_V$ definitions
can vary by $O(a)$.)
In Figs.~\ref{fig_zv+bv_zv_beta_magic_qu+dyn_zvpap}, 
\ref{fig_zv+bv_bv_beta_magic_qu+dyn_zvpap}
we plot our results for $Z_V$ and $b_V$ respectively for both quenched
and unquenched fermions. As, in particular for unquenched fermions the
available $\beta$-range is rather narrow and far from the continuum
limit the Pad{\'e} extrapolation should be treated with some caution.
However it is encouraging to note that the ordering of the points
is correct. Thus the Pad{\'e} results should be regarded mainly
as an interpolating formula between the various $\beta$ values.
Note however in the quenched case that the Pad{\'e} results
track quite well the ALPHA results.

Some of our earlier results for $Z_V$ for quenched fermions
can be found in \cite{capitani98a}. Note also an example of 
the large increase in the statistical fluctuations shown there
when matrix elements with moving nucleons ($\vec{p} \not= \vec{0}$)
are considered.


\section{$Z_V$ for unimproved fermions}

Although most of our results are for $O(a)$ improved fermions we
also have results at $\beta = 6.0$ for quenched unimproved fermions
($c_{sw} = 0$) which serve as a useful check on the improved results.
The method is identical to that described previously, so we shall just
give the results here. In Table~\ref{table_quenched_run_params_csw=0}
\begin{small}
\begin{table}[t]
   \begin{center}
      \begin{tabular}{||l|l|l|c||c||}
         \hline
\multicolumn{1}{||c}{$\beta$}  &
\multicolumn{1}{|c}{$\kappa$}  & 
\multicolumn{1}{|c}{$N_S^3\times N_T$}  &
\multicolumn{1}{|c||}{$\#$ configs.} &
\multicolumn{1}{c||}{$(\Delta R[V_4^{(u-d)}])^{-1}$}  \\
         \hline
6.0 & 0.1515 & $16^3\times 32$ & $O(980)$  & 0.7521(2) \\
6.0 & 0.1530 & $16^3\times 32$ & $O(1130)$ & 0.7220(2) \\
6.0 & 0.1550 & $16^3\times 32$ & $O(1360)$ & 0.6845(3) \\
6.0 & 0.1550 & $24^3\times 32$ & $O(220)$  & 0.6837(3) \\
6.0 & 0.1558 & $24^3\times 32$ & $O(220)$  & 0.6689(4) \\
6.0 & 0.1563 & $24^3\times 32$ & $O(220)$  & 0.6595(6) \\
6.0 & 0.1563 & $32^3\times 48$ & $O(250)$  & 0.6599(4) \\
6.0 & 0.1566 & $32^3\times 48$ & $O(410)$  & 0.6545(5) \\
         \hline
      \end{tabular}
   \end{center}
\caption{Parameter values used in the quenched simulations
         with unimproved fermions, together
         with $(\Delta R[V_4^{(u-d)}])^{-1}$.}
\label{table_quenched_run_params_csw=0}
\end{table}
\end{small}
we give the parameters of the runs and raw results. Note that
the range of quark masses is greater than in the $O(a)$ improved
case and that we have runs on different volumes at the same quark
mass. It is apparent from the table that finite volume effects are
rather small.

Plotting $(\Delta R[V_4^{(u-d)}])^{-1}$ against $am_q$
gives the result $Z_V = 0.6436(3)$ at $\beta = 6.0$
(for completeness the gradient is $0.9088(27)$),
using $\kappa_c = 0.157129(10)$.
We see that the numerical value is somewhat lower than the equivalent
$O(a)$ improved value. This result is consistent, but a little
larger than the recent determination given in \cite{aoki02a}
(although there different $\beta$-values are used). Again this is not
unexpected, as different determinations of $Z_V$ can now differ by 
$O(a)$ terms (see also \cite{capitani98a}).


\section{Conclusions}

Our method is not in disagreement with results
of other approaches for $O(a)$ improved quenched fermions.
For example even for $b_V$ there is only a few percent scatter in
comparison with other methods. This is what one would expect with $O(a)$ 
discrepancy between the various definitions.

$Z_V$ and $b_V$ are both further away from $1$ in the unquenched case
than in the quenched case at comparable lattice spacings (roughly
$a_{n_f=2}(5.25) \sim a_{n_f=0}(6.0)$). This is partially an effect
from the use of (numerically) smaller $\beta$ factors and also partially
because of the presence of fermions which induces a shift in the
effective $\beta$.
Finally we note that we cannot use one loop perturbation expansion results
in the present day quenched/unquenched $\beta$ regions, although this 
can be improved by using TI-RGI-BPT, eg \cite{capitani01a},
or BLM/TI methods, \cite{harada02a}.


\section*{Acknowledgements}

This work has been supported in part by
the European Community's Human Potential Program under contract
HPRN-CT-2000-00145, Hadrons/Lattice QCD,
and by the DFG (Forschergruppe Gitter-Hadronen-Ph\"anomenologie).

The numerical calculations were performed on the Hitachi {\it SR8000} at
LRZ (Munich), the Cray {\it T3E}s at EPCC (Edinburgh), NIC (J\"ulich) and
ZIB (Berlin) as well as the {\it APE100} and {\it APEmille} at NIC (Zeuthen).
The Edinburgh Cray {\it T3E} was supported by the UK Particle Physics and 
Astronomy Research Council under grants GR/L22744 and PPA/G/S/1998/00777.

We thank R. Sommer for useful comments.


\appendix

\section*{Appendix}

The result in eq.~(\ref{jump}) may alternatively be shown using
transfer matrix methods. As $J^{(q)}_4$ is conserved then
(with normalisation $\langle i | j \rangle = \delta_{ij}$)
\begin{equation}
   \langle i | J^{(q)}_4 | j \rangle = \chi_i^{(q)} \delta_{ij} \,,
\end{equation}
where $\chi_i^{(q)}$ is the `charge' (associated with the current
$J^{(q)}_\mu$) of the state $|i\rangle$. Thus inserting complete
sets of states into the three-point function gives
\begin{eqnarray}
   \langle B(t) J^{(q)}_4(\tau) \bar{B}(0) \rangle
          &=& \sum_{ij} \langle j|B|i\rangle \chi_i^{(q)}
                        \langle i|\bar{B}|j\rangle
                        e^{-aE_it}e^{-aE_j(N_T-t)}   \qquad 0 < \tau < t
                                                  \nonumber \\
          &=& \langle 0|B|N\rangle \chi^{(q)}_N
                \langle N|\bar{B}|0\rangle e^{-am_Nt} + \ldots \,,
\label{0letaulet}
\end{eqnarray}
(where the second equation assumes that the excited states
have died away and takes the charge on the nucleon, $\chi^{(q)}_N$, to be
$\chi^{(q)}_N = \langle N|J^{(q)}_4|N\rangle$) and similarly
\begin{eqnarray}
   \langle J^{(q)}_4(\tau) B(t) \bar{B}(0) \rangle
          &=& \sum_{ij} \chi^{(q)}_j 
                     \langle j|B|i\rangle \langle i|\bar{B}|j\rangle
                        e^{-aE_it}e^{-aE_j(N_T-t)} \qquad t < \tau < N_T
                                                  \nonumber \\
          &=&  \chi^{(q)}_{N^*}
               \langle \bar{N}^* |B|0\rangle 
               \langle 0|\bar{B}|\bar{N}^*\rangle e^{-am_{N^*}(N_T-t)} 
                            + \ldots \,,
\label{tletauleT}
\end{eqnarray}
$N^*$ being the parity partner of the nucleon and where
$\chi^{(q)}_{N^*} = \langle \bar{N}^* |J^{(q)}_4| \bar{N}^* \rangle
\equiv - \chi^{(q)}_N$. All of these expressions are independent of $\tau$.
The transfer matrix approach thus again predicts perfect plateaus
for $J^{(q)}_4$. Furthermore, the discontinuity is given by the difference
between these two expressions. As $\langle i|\bar{B}|j\rangle$ is only
non-zero if the difference in the charges of $|i\rangle$ and $|j\rangle$
is equal to that of the charge of a nucleon, then we can replace
$\chi^{(q)}_i - \chi^{(q)}_j$ by $\chi^{(q)}_N$ to give
\begin{eqnarray}
   \lefteqn{
     \langle B(t) J^{(q)}_4(\tau) \bar{B}(0) \left. 
        \rangle\right|_{0 < \tau < t} -
     \langle J^{(q)}_4(\tau) B(t) \bar{B}(0) \left. 
        \rangle\right|_{t < \tau < N_T} }
          & &                                     \nonumber \\
          &=&
              \sum_{ij} (\chi^{(q)}_i - \chi^{(q)}_j)
                \langle j|B|i\rangle \langle i|\bar{B}|j\rangle
                        e^{-aE_it}e^{-aE_j(N_T-t)}
                                                 \nonumber \\
          &=& \chi^{(q)}_N \langle B(t) \bar{B}(0) \rangle \,.
\end{eqnarray}
From eqs.~(\ref{0letaulet}), (\ref{tletauleT}) we see that
for $t < N_T/2$ (the case considered here) the exponential factor
is larger for the term $0 < \tau < t$ which means that $c^{(q)}_2$
is small in comparison with $c^{(q)}_1$.



\end{document}